\documentclass[cits]{PoS}
\title{Mapping the ionizing sources with CMB polarization measurements}
\ShortTitle{Mapping the ionizing sources with
CMB polarization measurements}

\author{ {L.A. Popa and C. Burigana}
\footnote{This work has been done in the framework of the
{\sc Planck} LFI activities.}\\

INAF/IASF Istituto Nazionale di Astrofisica,
Istituto di Astrofisica Spaziale e Fisica
Cosmica Bologna,
Via Gobetti 101, I-40129 Bologna, Italy\\
ISS Institute for Space Sciences, Bucharest-Magurele R-76900, Romania\\
E-mail: \email{popa@iasfbo.inaf.it burigana@iasfbo.inaf.it} }

\abstract{
With the three-year data, the {\it Wilkinson Microwave Anisotropy Probe}
(WMAP3) produced a more
accurate determination of the electron
scattering optical depth,
downwarding its value from $\tau_{es}=0.17\pm 0.08$
obtained with the first-year data (WMAP1)
to $\tau_{es}=0.09\pm 0.03$.
As a consequence,
the structure formation in the $\Lambda$CDM best fit model
obtained WMAP3 is delayed relative to that of WMAP1. \\
We show that
the delay of structure formation
can not fully account for the reduction of $\tau_{es}$
from WMAP1 to WMAP3 when the radiative transfer
effects and feedback mechanisms
are took into account in computing the reionization history
of the Universe. \\
We also show that a PopIII stellar cluster with
a mass of $80 M_{\odot}$ and a heavy Larson initial mass function
has an ionizing efficiency high enough to account for
WMAP3 results, while in the case of WMAP1,
a higher stellar mass of $1000M_{\odot}$ was required.\\
As the ultimate limit in constraining the reionizatin history
of the Universe with {\sc Planck}
will be placed by the our understanding of systematic effects
and foregrounds removal, we discuss also these aspects.}

\FullConference{
CMB and Physics of the Early Universe\\
		 20-22 April 2006\\
		 Ischia, Italy}


\begin{document}
\section{Introduction}

With the three-year data on the anisotropy of the cosmic microwave background (CMB) and its polarization,
the {\it Wilkinson Microwave Anisotropy Probe} (WMAP3) produced
a more accurate determination of the electron
scattering optical depth, downwarding its value from
$\tau_{es}=0.17\pm 0.08$ \cite{Spergel03}  obtained with the first-year data (WMAP1)
to $\tau_{es}=0.09\pm 0.03$ \cite{Page06, Spergel06},
consistent with an abrupt reionization at
redshift $z_{re} \simeq 11$, significantly later than $z_{re}\simeq 17$
as implied by WMPA1.

Other most important changes of the cosmological parameters
from WMAP1 to WMAP3 are the reduction of the normalization of
the power spectrum on large scales ($\sigma_8=0.92\rightarrow0.76$)
and the reduction of the scalar spectral index of the primordial density
perturbations ($n_s=0.98\rightarrow0.74$).
As a consequence,
the structure formation in the $\Lambda$CDM model with
the primordial power spectrum of the density fluctuation obtained by WMAP3
is delayed relative to that of WMAP1.\\
Based on the simple assumption of constant ionizing efficiency,
a recent paper \cite{alvarez06} claims that the delay of structure formation
controls the reionization in WMAP3 best fit model such that,
if ionizing efficiency is large enough to make reionization early
and $\tau_{es}=0.17$ in WMAP1 case, the same efficiency is required
to have the reionization later and $\tau_{es}=0.09$
in WMAP3 case.

In this paper we show that
the delay of structure formation
can not fully account for the reduction of $\tau_{es}$ value
from WMAP1 to WMAP3 best fit models when the radiative transfer
effects and feedback mechanisms
are take into account in computing the reionization history
of the Universe. We also show that a PopIII stellar cluster with
a mass of $\sim 80 M_{\odot}$ and a heavy Larson IMF has
an ionizing efficiency high enough to account for the
or the value of $\tau_{es}$ obtained by WMAP3.

{\sc Planck} surveyor will have enough sensitivity
the test various reionization models even when
they imply the same value for $\tau_{es}$ \cite{PBM05}.\\
As the ultimate limit in constraining the reionizatin history
of the Universe with {\sc Planck}
will be placed by the our understanding of systematic effects
and foregrounds removal, we discuss also these aspects.

\section{WMAP 3-year data: implications for
the properties of ionizing sources}
We compute the reionization histories of the Universe
for the emission of a PopIII stellar cluster of mass $M$ and
a heavy Larson IMF
for different values of the parameters $(n_s,\sigma_8,M)$.
Our computation includes all the radiative mechanisms
relevant for the primordial
gas dynamics: photo-ionization, photo-heating  and cooling of the
hydrogen and helium in the expanding Universe.
The mean UVB flux is obtained as  solution
to the radiative transfer equation
by assuming  a constant star formation efficiency $f_*=0.1$.
The details of the
computation can be found in \cite{Popa06}.\\
The model parameters and the corresponding values of $\tau_{es}$ are given in Table~1.
The model WMAP* was constructed to have the same values for $n_s$ and $\sigma_8$ as
WMAP1 and the same stellar mass as WMAP3.

We find that a stellar cluster with a mass of $M\simeq 80M_{\odot}$
has an ionizing efficiency high enough to account
for  WMAP3 value of $\tau_{es}$
while for the case of WMAP1 a higher stellar mass,
$M\simeq 1000M_{\odot}$, is needed.
For WMAP* model we obtain a value of the electron optical
depth of $\simeq0.13$ that can account for about $80\%$ from that obtained by WMAP1.

On the basis of these calculations,
we conclude that, althrough WMAP3 has not enough sensitivity
to constrain the reionization history \cite{Spergel06},
the delay of structure formation
in WMAP3 best fit model can not fully account for
the reduction of $\tau_{es}$ from WMAP1 to WMAP3 when
the radiative effects and feedback mechanisms  are take into account.
\begin{table}[]
\begin{small}
\begin{center}
\begin{tabular}{cccccc}
\hline \hline
Model  & $n_s$&$\sigma_8$& $M/M_{\odot}$ & $\tau_{es}$   \\ \hline
WMAP1  & 0.99$\pm$0.04 & 0.92$\pm$0.1& $1000$ & 0.157$\pm$0.032 \\
WMAP3  & 0.961$\pm$0.017 &0.76$\pm$0.05&  $80$      & 0.093$\pm$0.012 \\
WMAP*& 0.99$\pm$0.04& 0.92$\pm$0.1&  $80$ & 0.130$\pm$0.032 \\ \hline
\end{tabular}
\end{center}
\end{small}
\caption[] { Model parameters}
\end{table}

\section{Perspectives from the {\sc Planck} mission: sensitivity and systematic effect control}
A fundamental aspect in constraining the reionization history
with {\sc Planck} CMB measurements,
which is necessary to fully exploit its increasing sensitivity,
is the accurate control of all systematic effects, of instrumental
and astrophysical origin and of their interplay.\\
From the point of view of the angular power spectrum recovery,
an useful classification of systematic effects can be based on their
different relevance at different multipoles.\\
Sidelobe pickup from Galactic emission and CMB dipole
\cite{Burigana04, Gruppuso06}
will mainly affect the low multipole region
from $\sim$ the dipole scale to
$\sim$ the first acoustic peak
(in the possible presence of bad optical behaviours and depending also
on the scanning strategy the straylight from inner Solar System
bodies could also affect the data).
These effects can be accurately removed (for example thorugh iterative methods)
during data analysis only in the presence of a very accurate understanding of the
instrument optical properties in work conditions.\\
Main beam distortions \cite{Burigana98, Mandolesi00}
mainly affect the highest multipole region accessible
to the resolution of a given receiver, i.e. the multipoles region
of secondary acoustic peaks. Again, the removal of this effect
based on the accurate evaluation of the effective window function
(because of the coupling between beam window function and scanning strategy)
or on deconvolution codes requires a very accurate knowledge of the main
beam shape, possible in flight using the transit on the main beam
of external planets (or, in general, of bright, stable,
and point-like sources).\\
Pointing errors could also affect particularly the high multipole region,
but this problem has been largely reduced in the {\sc Planck} mission
by the use of high precision (up to few arcseconds)
star trackers. \\
Non-idealities in LFI radiometers
and bolometers introduce  $1/f$-like noise while long term drifts
can be also induced by temperature instability
\cite{Seiffert02,Mennella02}.
These effects have impact at both low and high multipoles because of the
stripes induced in the maps. Dedicated destriping and map-making methods
have been implemented and tested to reduce their effect in the map
and in the angular power spectrum.
%Delabroui98,
\cite{Maino99, Gasperis05}.
Only a largely reduced excess of power at
low multipoles (because of noise long term correlation) is the most remarkable
effect remaining after data reduction, that can be properly modelled
through simulations.\\
Since LFI and HFI could in principle have different responses
to the various systematic effects,
a powerful tool for the detection and reduction
of systematic effects in the {\sc Planck} data
will derive also from the accurate comparison
between the data at the closest frequencies of LFI and HFI,
respectively the 70~GHz and 100~GHz channels, where also
foreground contamination is expected to be almost minimal.\\
In general, the {\sc Planck} goal is the achieve a suppression
of any systematic effect
at a level better than $\simeq 3 \mu$K ($\simeq 1 \mu$K)
in terms of peak-to-peak ({\sc rms})
spurious signal.\\
The control of systematic effects and the wide frequency
coverage of {\sc Planck}, allowing an accurate modelling and
removal of foregrounds, will make this mission
almost limited only by cosmic variance, at least for
total power data.
For polarization measurements, the situation is probably
more complicated because of the intrinsic weakness of CMB
polarization anisotropy, the increasing of complexity
in systematic effect control and subtraction, and finally,
the largest relative weight of polarized foregrounds
that need to be understood and removed at few percent accuracy level,
or better, for an accurate study of CMB polarization anisotropy.\\
On the other hand, at least for the $E$ mode, {\sc Planck}
is expected to reach an accuracy level
sufficient not only to precisely
measure the Thomson scattering optical depth but also
to constrain among various possible
cosmological reionization histories \cite{PBM05}.


\begin{thebibliography}{99}
\bibitem{Spergel03}
D.N.~Spergel et al., \emph{First-Year Wilkinson Microwave Anisotropy Probe (WMAP) Observations:
Determination of Cosmological Parameters},
\emph{ApJs} {\bf 148} (175) 003
\bibitem{Page06}
L.~Page, et al., \emph{Three Year Wilkinson Microwave Anisotropy Probe (WMAP) Observations:
Polarization Analysis},
 \emph{ApJ} (submitted) 006 [astro-ph/0603450]
\bibitem{Spergel06}
D.N.~Spergel et al.,
\emph{Wilkinson Microwave Anisotropy Probe (WMAP) Three Year Results: Implications for Cosmology},
 \emph{ApJ} (submitted) 006 [astro-ph/0603449]
\bibitem{alvarez06}
M.A.~Alvarez, P.R.~Shapiro, K.~Ahn, I.T.~Iliev,
\emph{Implications of WMAP Three Year Data for Reionization},
006 [astro-ph/0604447]
\bibitem{Popa06}
L.A.~Popa, \emph{WMAP 3-year polarization data: Implications for the reionization history },
006  [astro-ph/0605358]
\bibitem{PBM05}
L.A.~Popa, C. Burigana, N. Mandolesi
\emph{ Radiative effects by high-z UV radiation background:
Implications for the future CMB polarization measurements},
\emph{N.A.} {\bf 147} (175) 005
\bibitem{Burigana04}
C. Burigana et al.,
\emph{Trade-off between angular resolution and straylight contamination in the {\sc Planck} Low Frequency Instrument. II. Straylight evaluation}
\emph{A\&A} {\bf 428}, 311, 04
\bibitem{Gruppuso06}
A. Gruppuso,  C. Burigana, F. Finelli,
\emph{Dipole straylight contamination and low multipoles},
(this conference)
\bibitem{Burigana98}
C. Burigana et al.,  \emph{Beam distortion effects on anisotropy measurements of the cosmic microwave background}
\emph{A \&As} {\bf 130}, 551, 998
\bibitem{Mandolesi00}
N. Mandolesi et al.,  \emph{On the performance of Planck-like telescopes versus mirror aperture}, \emph{A \& As} {\bf 145}, 323, 000
\bibitem{Seiffert02}
M.  Seiffert et al., \emph{1/f noise and other systematic effects in the Planck-LFI radiometers} \emph{A\& A} {\bf 391}, 1185, 002
\bibitem{Mennella02}
A. Mennella et al., \emph{PLANCK: Systematic effects induced by periodic fluctuations of arbitrary shape}, \emph{A\&A} {\bf 384}, 736, 002
\bibitem{Maino99}
D. Maino, \emph{The Planck-LFI instrument: Analysis of the 1/f noise and implications for the scanning strategy},
\emph{A\&As} {\bf 140}, 383, 999
\bibitem{Gasperis05}
G. de Gasperis et al.,\emph{ROMA: A map-making algorithm for polarised CMB data sets}, \emph{A \&A} {\bf 436}, 1159, 005
\end{thebibliography}
\end{document}